\documentclass[seceq]{ptptex}

\usepackage{graphicx}
\usepackage{wrapft}

\newcommand{\beq}{\begin{equation}}
\newcommand{\beqa}{\begin{eqnarray}}
\newcommand{\eeq}{\end{equation}}
\newcommand{\eeqa}{\end{eqnarray}}

\newcommand{\lsim}{\lesssim}
\newcommand{\gsim}{\gtrsim}

\title{Determination of the equation of the state of the  Universe
 using $\sim 0.1$ Hz Gravitational Wave Detectors  
}

\author{Ryuichi \textsc{Takahashi}$^{1}$ and Takashi
 \textsc{Nakamura}$^{2}$
}

\inst{$^1$Division of Theoretical Astrophysics,
 National Astronomical Observatory, Mitaka, Tokyo 181-8588, Japan \\
$^2$Department of Physics, Kyoto University, Kyoto 606-8502, Japan
}

\abst{
We show that ten(one) years operation of the ultimate DECIGO
(DECihertz Interferometer Gravitational wave Observatory)
can determine the cosmic equation of the state with such accuracy
 that 0.06\%(3\%), 0.08\%(4\%) and 0.06\%(3\%) for $\Omega_m$, $\Omega_w$
 and $w$, respectively. 
In more realistic case of practical DECIGO or BBO (Big Bang Observer),
 $w$ will be determined within $\sim 10\%$ by ten years observation
 assuming the flat universe model.     
Hence, the DECIGO or BBO will give an independent determination
 of the cosmic equation of the state. 
}

\begin{document}

\maketitle

\section{Introduction}

The laser interferometers begin to search for the astrophysical
 gravitational wave sources in broad frequency band. 
For the higher frequency band of  $10-10^3$ Hz, the ground-based
 interferometers such as LIGO, TAMA, VIRGO, GEO are currently in
 operation, and the next generation of interferometers such as
 advanced LIGO and LCGT \cite{lcgt} are planed.
For the lower frequency band of $10^{-4}-10^{-1}$ Hz, the
 spaced-based interferometer LISA (the Laser Interferometer
 Space Antenna) \cite{lisa} will be launched in $\sim 2012$.
To fill the frequency gap between the ground-based detectors
 and LISA of $10^{-2}-10$ Hz, the deci hertz laser interferometer
 in space is planed to be constructed in $\sim 2020$.
This detector is called DECIGO (DECihertz Interferometer Gravitational
 wave Observatory) in Japanese group \cite{skn01,tn03} or BBO
 (Big Bang Observer) in the NASA SEU 2003 Roadmap "Beyond Einstein"
 \cite{bbo}.

The major scientific objectives of the decihertz antenna are:

\begin{itemize}
\item The primordial gravitational wave background will be observed,
 since there are no or little confusion noise from galactic white
 dwarf binaries around $0.1$ Hz \cite{uv01,skn01}. 
\item $\sim 10^5$ chirp signals of coalescing binary neutron stars
 and stellar mass black holes per year will be detected.
 By analysing the signals from these binaries at cosmological
 distances, it may be possible to determine the expansion rate of the
 Universe and the equation of state for the dark energy \cite{skn01}.
\item The merger of intermediate mass black holes $(10^2-10^5 M_{\odot})$
 will be observed and hence it is helpful to understand the formation and
 growth history of super massive black holes. \cite{b04} 
\end{itemize} 
In this paper, we investigate the second objective, i.e. the
 determination of the expansion rate of the Universe by the
 decihertz gravitational wave detector.

Recently, from the observation of distant supernovae the expansion
 of the Universe appears to be accelerating at  present,
 which provides the evidence for a non zero cosmological constant or
 more generally the dark energy with the negative pressure
 (see Ref.\citen{r04} and references therein).
In near future, SNAP (the Supernova/Acceleration Probe) will
 observe $\sim 2000$ supernovae per year for $z < 1.7$ and
 will determine the energy density
 of the dark energy and its equation of the state with high accuracy
 (within a few \% error) \cite{snap04}.   
But the measurement of supernovae might have possible systematic
 uncertainties such as the dust extinction, the gravitational lensing
 and the metallicity dependence.
Hence it is desirable to determine the nature of the dark energy by an
 independent method, since the nature of the dark energy is by far the
 most important to clarify its origin theoretically.

Seto, Kawamura and Nakamura (2001) suggested that the acceleration rate
 of the Universe will be directly determined by a $10$ yr observation
 of the gravitational wave from a neutron star binary at $z=1$. 
This is because the phase of the signal from the distant source is
 modulated due to the cosmic acceleration and this modulation can be
 measured by using the matched filtering techniques.
The host galaxy of the source can be identified by the
 follow-up observations
 using electromagnetic wave antennae since the angular
 resolution of the ultimate DECIGO is a few arcsecond
 even for the source with $z\sim 1$, and hence the 
 source redshift is also determined.\cite{tn03}\footnote{In Ref.~\citen{tn03},
 the source is at 300Mpc but the sensitivity is
  $\sim$ 1000 times worse than the ultimate DECIGO. It is possible to predict
  the source position within an arcminute in Ref.~\cite{tn03} so that for
  the source at $\sim 10$Gpc ($z \sim 1$) the ultimate DECIGO can determine
  the position within a few arcsecond.}
Thus the expansion rate of the Universe (i.e. Hubble parameter $H(z)$)
 is directly obtained at the source redshift $z$.

Following Seto et al.(2001), we evaluate the accuracy of
 determining the
 cosmic acceleration rate by the decihertz gravitational wave antenna.
We newly include the effects of (1) the number of  NS/NS merger events
 and (2) the various masses and redshifts of the binaries.
From the 1st effect, since $\sim 10^5$ signals of NS/NS binaries will 
 be measured per year, the accuracy of the determination of the
 cosmic acceleration is $1/\sqrt{10^5} \sim 10^{-2}-10^{-3}$ times
 better than that for a single event.
From the 2nd effect, we investigate the dependence of the accuracy
 on the binary masses $(0.1-10^5 M_{\odot})$ and its redshifts
 $(z=0-2)$.

We adopt $(\Omega_m,\Omega_{\Lambda})=(0.3,0.7)$
 cosmology with the Hubble parameter $H_0=70 \text{km}
 \text{s}^{-1} \text{Mpc}^{-1}$ and use the units of $c=G=1$.

\section{Basics}

\subsection{Sensitivity of DECIGO/BBO}

\begin{wrapfigure}{l}{6.6cm}
\centerline{\includegraphics[width=6.4cm,height=6.4cm]{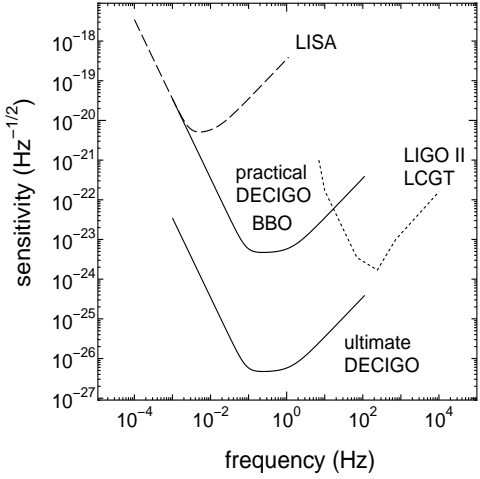}}
\caption{Sensitivity of DECIGO/BBO.}
\label{fnc3}
\end{wrapfigure}

DECIGO/BBO would consist of three spacecrafts orbiting the sun, in a 
 triangular configuration with separations of $5 \times 10^4$ km,
 which is $100$ times smaller than that of LISA \cite{bbo}.
The strain sensitivity of the practical DECIGO or BBO is about $1000$ times
 better than that of LISA (i.e., $10^{-23} {\mbox{Hz}}^{-1/2}$),
 and acceleration noise is $100$ times smaller than that of LISA
 (i.e. $3 \times 10^{-17} {\mbox{m}} ~{\mbox{s}}^{-2} ~{\mbox{Hz}}^{-1/2}$).
Using the above assumptions and the sensitivity curve of LISA in
 Ref.\citen{bc04},
 we plot the sensitivity of practical DECIGO and BBO in Fig.\ref{fnc3}
 \cite{lhh00}.
We also show that of LIGO II (or LCGT).
The sensitivity of the ultimate DECIGO is $1000$ times better than
 that of the practical DECIGO (i.e. $h \sim 10^{-27}$ at
 $f=0.1$ Hz), which is determined by the quantum limit sensitivity
 for a $100$ kg mass mirror \cite{skn01}.
In this paper, we use the sensitivity curve of ultimate DECIGO, to show what we can do ultimately.

\subsection{Gravitational Waveform of In-spiralling Binary in
 Accelerating Universe}

We consider a binary of mass $M_{1,2}$ at the redshift $z$ 
 as the gravitational wave source.
As the binary emits the gravitational radiation, its orbital
 separation decreases to zero.  
From the Newtonian quadrupole approximation \cite{cf94}, 
 the time left till the coalescence is given by, 
 $\Delta t \equiv t_c - t = 5 (8 \pi f)^{-8/3} \mathcal{M}_z^{-5/3}$,
 where $t_c$ is a coalescence time, $f$ is the frequency of
 the gravitational waves, and $\mathcal{M}_z =
 (M_1 M_2)^{3/5} (M_1 + M_2)^{-1/5} (1 + z)$ is a redshifted chirp mass.

The time to the coalescence $\Delta t$ at present is related to that
 ($\Delta t_z$) at the redshift $z$ as \cite{skn01,loeb98},
\beqa
  \Delta T = \Delta t - X(z) \Delta t^2, 
\label{delt}
\eeqa
where $\Delta T=\Delta t_z (1+z)$ and the second term is the correction
 term due to the cosmic acceleration.
The acceleration parameter $X(z)$ is defined by \cite{skn01},
\beq
 X(z) = \frac{1}{2} \left[ H_0- \frac{H(z)}{1+z} \right],
\label{xz} 
\eeq
where $H(z)$ is the Hubble parameter at $z$. 
$X$ is rewritten as $X=(\dot{a}(0)-\dot{a}(z))/2$ where $a$ is a 
 scale factor and $\dot{ } =d/dt$.
Thus if $X(z)>0$ $[X(z)<0]$, $\dot{a}(0) > \dot{a}(z)$
 $[\dot{a}(0) < \dot{a}(z)]$ and the universe is accelerating
 (decelerating) (see Ref.\citen{skn01}).
For the small redshift $z \ll 1$, $X(z)$ is simply reduced to
 $X(z)/H_0=-(1/2) q_0 z$ where $q_0$ is the deceleration parameter.
The phase of the gravitational wave signal at present 
 is shifted by $- 2 \pi f X(z) \Delta t^2$ from Eq.(\ref{delt}).

In the frequency domain, the gravitational waveform including the
 effects of the cosmic acceleration is written as \cite{skn01},
\beq
  \tilde{h}_{\text{acc}}(f) = \tilde{h}(f) e^{i \delta
 \Psi_{\text{acc}}(f)},
\label{hacc}
\eeq
where the phase correction is $\delta \Psi_{\text{acc}} $
 $= - 2 \pi f X(z) \Delta t^2 $ or,
\beq
 \delta \Psi_{\text{acc}} = - \frac{25 \pi^{-13/3}}{32768} X(z) f^{-13/3}
 \mathcal{M}_z^{-10/3}.
\label{psiacc}
\eeq
$\tilde{h}(f)$ in Eq.(\ref{hacc}) is the in-spiral waveform and we 
 use the restricted first post-Newtonian approximation.
Then, we have $\tilde{h}(f) = A f^{-7/6} e^{i \Psi(f)}$,
where $A$ is the amplitude and $\Psi(f)$ is the phase.\cite{cf94}
 They depend
 on five parameters: the redshifted chirp mass $\mathcal{M}_z$, the
 reduced mass $\mu_z = M_1 M_2 (1 + z)/(M_1 + M_2)$, 
 the coalescence time $t_c$, the phase $\phi_c$, and the luminosity distance
 to the source $D_S$. The amplitude is given by $A = K (5/96)^{1/2}
 \mathcal{M}^{5/6} /(\pi^{2/3} D_S)$, where $K$ is the constant determined
 by the inclination of the source, the relative orientation of the source,
 and the detector. Since the average value of $K$ is about
 unity \cite{fc93}, we assume $K$ = 1 for the following calculation.
 The phase $\Psi(f)$ is a rather complicated function of
 $\mathcal{M}_z$, $\mu_z$, $\phi_c$, and $t_c$, which is given in
 Eq.(3.13) of Cutler \& Flanagan (1994).

\subsection{Parameter Estimation}

The signal $\tilde{h}_{acc} (f)$ in Eq.(\ref{hacc}) is characterized by
 six parameters $\gamma_i =$ ($\mathcal{M}_z$, $\mu_z$, $\phi_c$,
 $t_c$, $D_S$, $X$).
In the matched filter analysis with the template, these parameters
 can be determined.
We compute the errors in the estimation of these parameters using the
 Fisher matrix formalism \cite{finn92,cf94}
 (see also Refs.\citen{cut98,hug02,mh02,seto02,vec03}).
The variance-covariance matrix of the
 parameter estimation error $\Delta \gamma_i$ is given
 by the inverse of the
 Fisher information matrix $\Gamma_{ij}$ as
 $\langle \Delta  \gamma_i \Delta \gamma_j \rangle
 = \left( \Gamma^{-1} \right)_{ij}$.
The Fisher matrix becomes
\beq 
  \Gamma _{ij}
 = 4 {\mbox{Re}} \int \frac{df}{Sn(f)}~
 \frac{\partial \tilde{h}_{\text{acc}}^{*}(f)}{\partial \gamma_i}
 \frac{\partial \tilde{h}_{\text{acc}}(f)}{\partial \gamma_j},
\label{fis}
\eeq
where $Sn(f)$ is the noise spectrum.
We regard $Sn(f)$ as the instrumental noise of ultimate DECIGO
 in Fig.\ref{fnc3},
 neglecting the binary confusion noise since there is no
 or little confusion noise for $f \gsim 0.1$ Hz \cite{skn01}.
The signal to noise ratio ($S/N$) is given by
\beq
  (S/N)^{~2} = 4 \int \frac{df}{Sn(f)}~
 \left| \tilde{h}_{acc}(f) \right|^2.
\label{snr}
\eeq
We integrate the gravitational waveform
 in Eq.(\ref{fis}) and Eq.(\ref{snr}) from $1$, $3$,
and  $10$ yr before the final merging to the
 cut-off frequency $f_{cut}$ when the binary separation
 becomes $r=6 (M_1 + M_2)$. 

The estimation error $\Delta \gamma$  simply scales as $(S/N)^{-1}$
 from Eq.(\ref{fis}) and Eq.(\ref{snr}).
Hence the errors for the practical DECIGO or BBO is simply $1000$ times
 larger than that for the ultimate DECIGO (see Fig.\ref{fnc3}).

\subsection{Event Rate of NS/NS Coalescence}

The estimation error $\Delta \gamma$ discussed in the previous section
 is for a single event.
But for a number of  events the estimation error decreases
in proportion to  (the number of
 events$)^{-1/2}$. 
For this purpose we first estimate the event rate of NS/NS coalescence.  

The NS/NS merger rate between the redshift $(z,z+dz)$ is written
 as,
\beq
  \frac{dN}{dz} = 4 \pi n_0 \frac{r(z)^2}{H(z)}
\eeq
where $r(z)$ is the comoving distance and $n_0$ is the comoving
 number density of the NS/NS merger rate per year.
Here, we assume that the comoving merger rate $n_0$ is constant. 
We choose $n_0$ so that $1$ event per year could be observed within
 $60$ Mpc from recent studies \cite{bur03,kal04},
 and hence we set $n_0=1 \times 10^{-6}
 \textrm{Mpc}^{-3} \textrm{yr}^{-1}$.  
\begin{wrapfigure}{l}{6.6cm}
\centerline{\includegraphics[width=6.4cm,height=6.4cm]{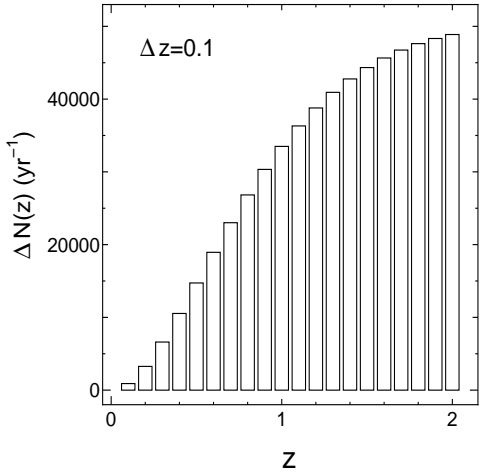}}
\caption{The NS/NS merger rate per year for $\Delta z=0.1$ bin.}
\label{fdn2}
\end{wrapfigure}
In Fig.\ref{fdn2}, the merger rate per redshift bin $\Delta z=0.1$,
 $\Delta N \equiv (dN/dz) \Delta z$, is shown.
Since the ultimate DECIGO could observe the NS/NS binaries at
 cosmological distances ($z \sim 1$) with high S/N $\sim 10^4$, 
 the NS/NS merger will be observed $10^{4-5}$
 events per year from Fig.\ref{fdn2}.
Thus, the estimation error $\Delta \gamma$ is $1/\sqrt{\Delta N}$
 $(\sim 10^{-2} ~{\text{or}}~ 10^{-3})$ times better than that
 for a single event.

\section{Results}

\begin{figure}[t]
\parbox{\halftext}{
  \includegraphics[width=6.4cm,height=6.4cm]{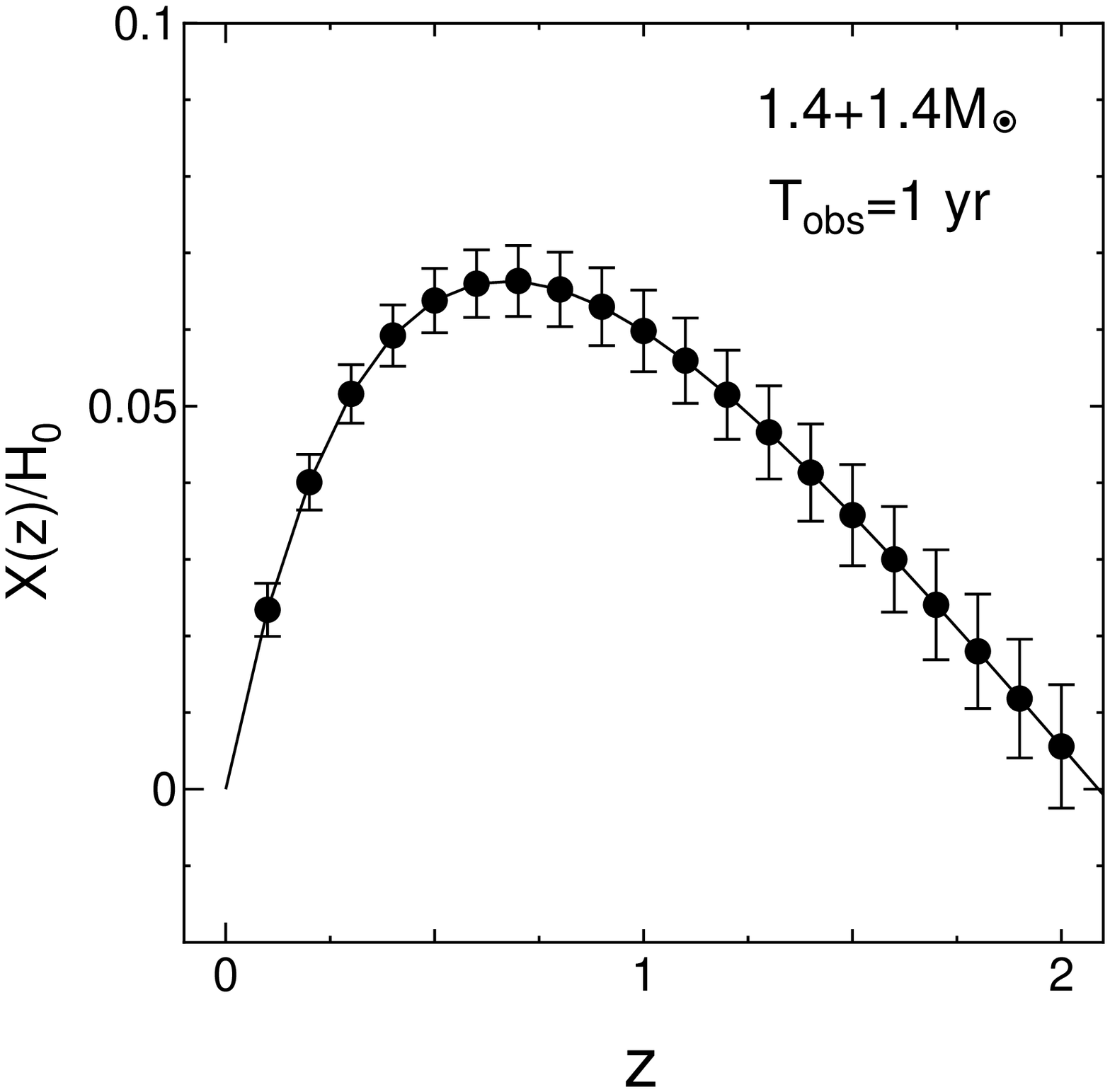}
\caption{The acceleration parameter $X(z)$ (in Eq.(\ref{xz})) as a
 function of redshift. The solid line is the case of $\Lambda$CDM
 cosmology ($h=0.7, \Omega_m=1-\Omega_\Lambda=0.3$).
The filled circles and the error bars are the mean values of $X(z)$
 and the estimation errors $(\Delta X/H_0)/(\Delta N)^{1/2}$ for NS/NS
 binaries $(M_{1,2}=1.4M_\odot)$ in $\Delta z=0.1$ bins.
We evaluate for $1$ yr observation before coalescence by the ultimate DECIGO.}
\label{dx18p}}
\hfill
\parbox{\halftext}{\vspace{-2.22cm}
  \includegraphics[width=6.4cm,height=6.4cm]{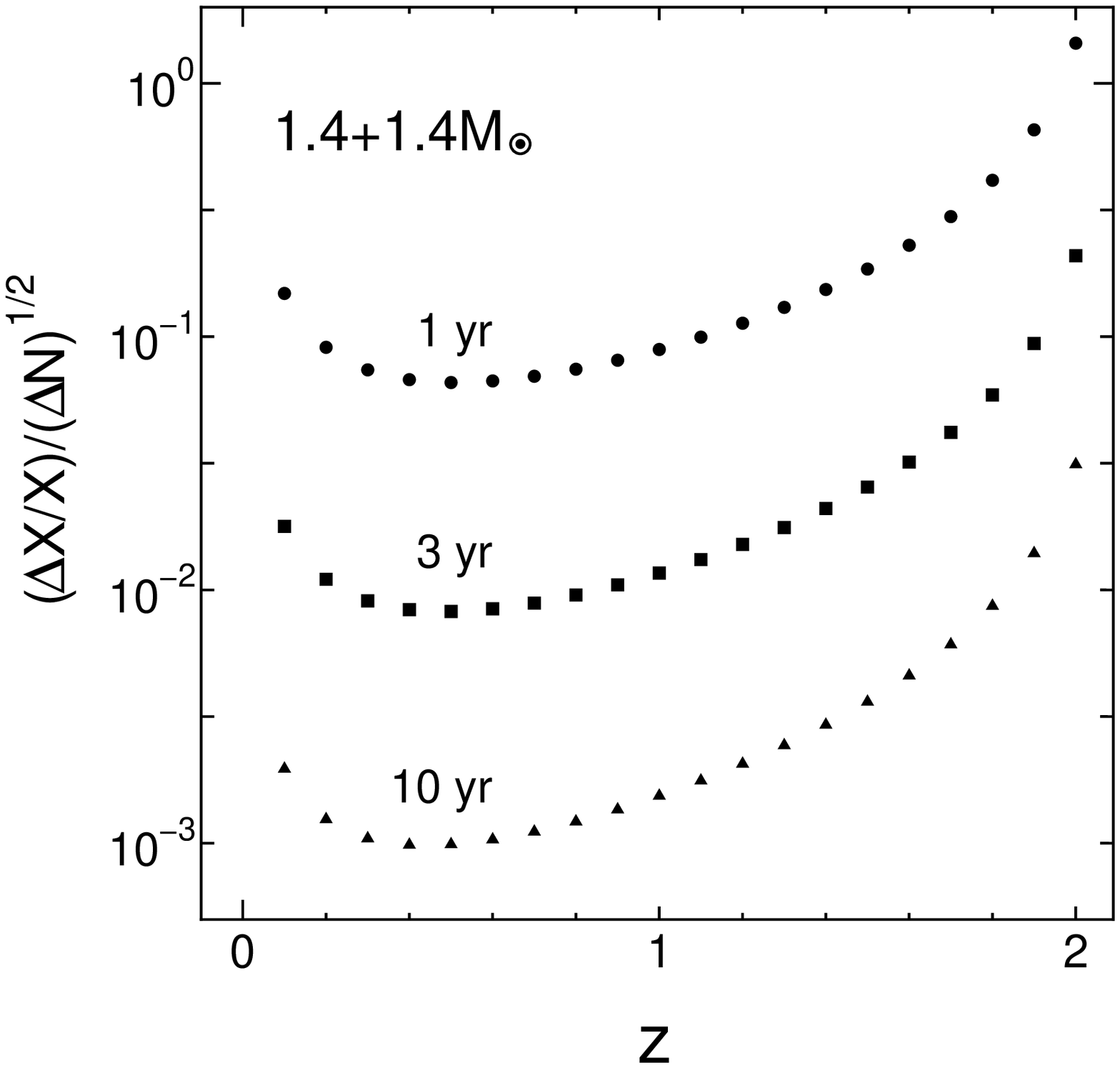}
\caption{Similar to Fig.\ref{dx18p}, but
 the relative estimation errors $(\Delta X/X)/(\Delta N)^{1/2}$
 are shown for various observational period 1,3 and 10 yr.}
\label{dx21p}}
\end{figure}

We show the estimation error of the cosmic acceleration parameter
 $X(z)$ in Eq.(\ref{xz}) at $z=0 - 2$.
We consider the mergers of NS/NS binaries $(M_{1,2}=1.4 M_{\odot})$
 as the sources detected by the ultimate DECIGO.
We assume $1$ yr observation before the coalescence.
We find that the signal-to-noise ratio is $\sim 2 \times 10^4
 (H_0 D_S)^{-1}$, where $D_S$ is the luminosity distance to
 the source.
Fig.\ref{dx18p} is the acceleration parameter $X(z)$ in Eq.(\ref{xz})
 as a function of $z$.
The solid line is the case of $\Lambda$CDM cosmology
 ($h=0.7, \Omega_m=1-\Omega_{\Lambda}=0.3$). 
The filled circles and the error bars are the mean values of $X(z)$ and  the estimation errors
 $(\Delta X/H_0)/(\Delta N)^{1/2}$ in $\Delta z=0.1$ bins.
Thus, the cosmic acceleration rate at $z=0-2$ will be directly determined
 by $1$ yr observation of the ultimate DECIGO.
We note that the Hubble parameter $H(z)$ is also measured 
 by using Eq.(\ref{xz}) : $H(z)=(1+z) [H_0-2 X(z)]$ for
 the given $H_0$ by another method.

Fig.\ref{dx21p} is similar to Fig.\ref{dx18p}, but
 the relative estimation errors $(\Delta X/X)/(\Delta N)^{1/2}$
 are shown for various observational period 1,3 and 10 yr.
The errors are the smallest at $z \sim 0.5$ because $X(z)$ is the largest there (see Fig.\ref{dx18p}).
The errors are $\approx 0.1~(10^{-3})$ for $1~(10)$ yr observation.
In the matched filtering analysis, the phase modulation $\delta
 \Psi_{\text{acc}}$ in Eq.(\ref{psiacc}) is roughly measured to an
 accuracy $\approx (S/N)^{-1}$. 
Hence the relative error $(\Delta X/X)/(\Delta N)^{1/2}$ is estimated as
\beq
  \frac{\Delta X}{X} \frac{1}{\sqrt{\Delta N}} \approx 0.02 
 \left( \frac{X/H_0}{0.05} \right)^{-1}
 \left( \frac{f}{0.1 {\text{Hz}}} \right)^{-1}
 \left( \frac{\Delta t}{1 {\text{yr}}} \right)^{-2} 
 \left( \frac{S/N}{10^4} \right)^{-1}
 \left( \frac{\Delta N}{10^4} \right)^{-1/2}.
\label{delx}
\eeq
The above estimation is roughly consistent with the results in
 Fig.\ref{dx21p}.
The errors are proportional to $\Delta t^{-2}$
 as shown in Eq.(\ref{delx}).
For the high redshift $z \gsim 1$  the error is  large, since
 both $S/N$ and $X(z)$ become small.

\begin{figure}[htb]
\parbox{\halftext}{
  \centerline{\includegraphics[width=6.4cm,height=6.4cm]{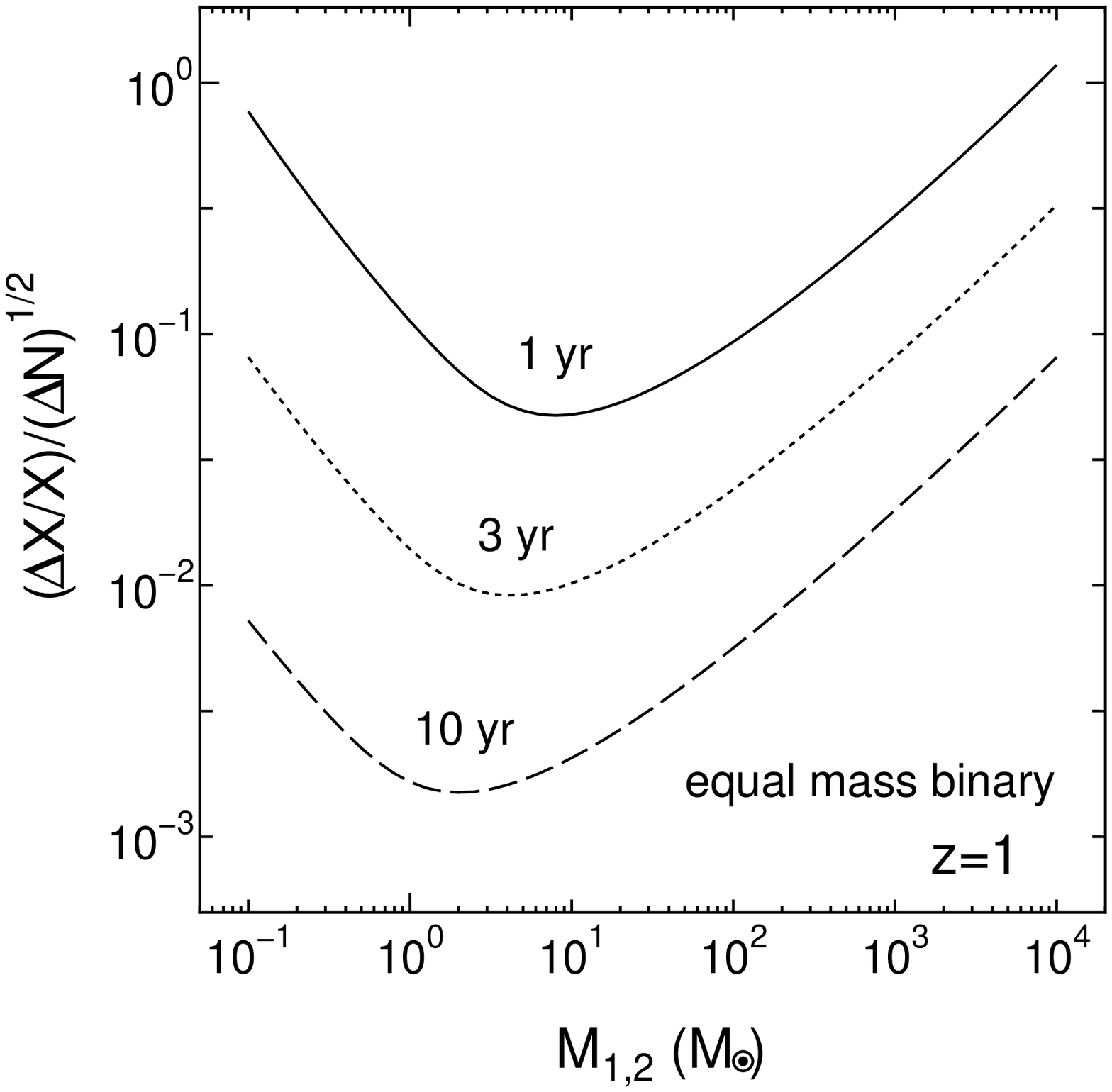}}}
\hfill
\parbox{\halftext}{
  \centerline{\includegraphics[width=6.4cm,height=6.4cm]{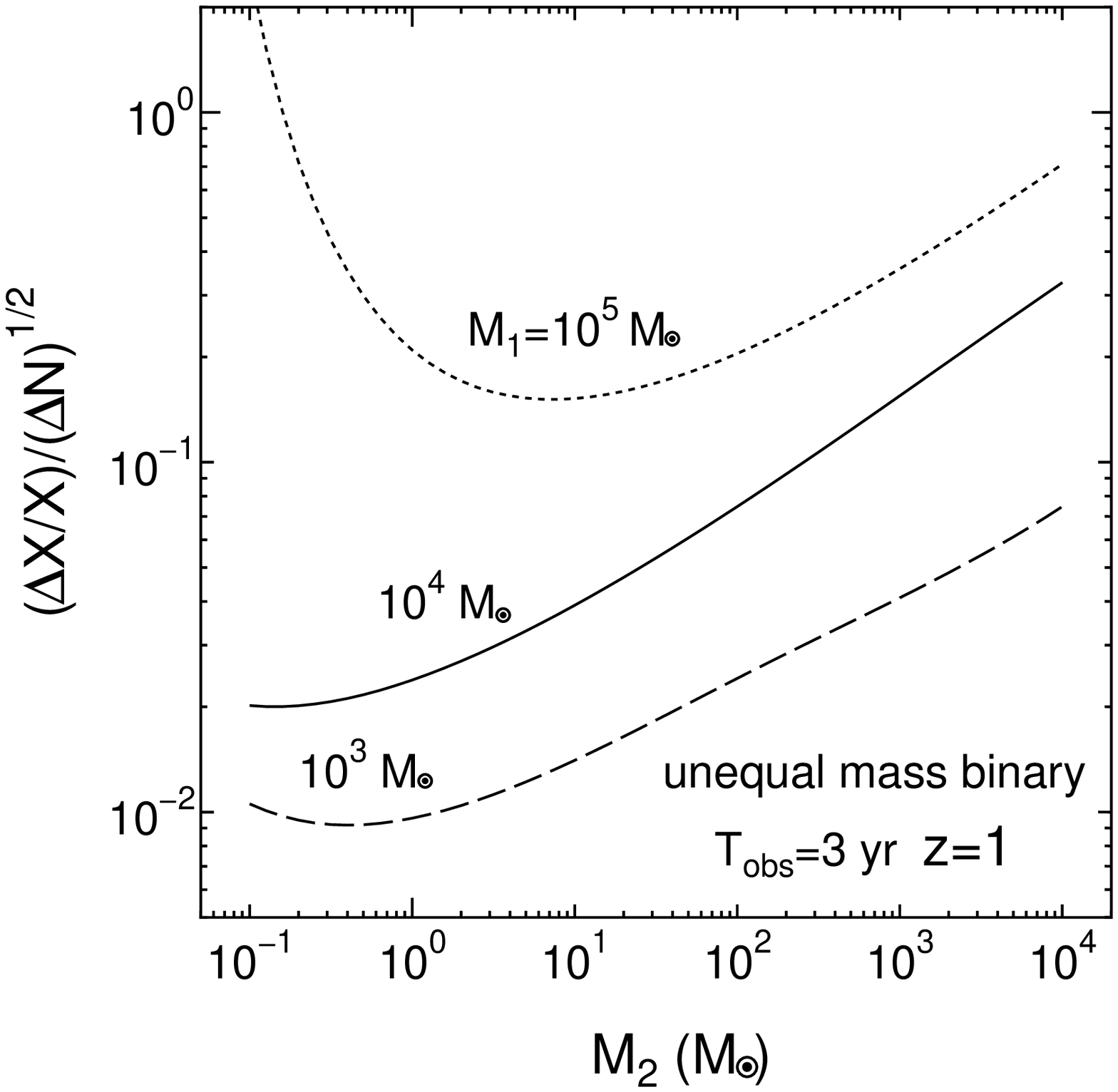}}}
\caption{Same as Fig.\ref{dx21p}, but as a function
 of the binary mass with the fixed redshift $z=1$.
The left panel is shown for the equal mass binary, and the right panel is 
 shown for the unequal mass binary ($M_1=10^3,10^4$ and $10^5 M_\odot$).
In right panel, the observational period is fixed with $3$ yr.  
}
\label{dx26p}
\end{figure}

Fig.\ref{dx26p} is the same as Fig.\ref{dx21p}, but as a function
 of the binary mass with the fixed redshift of $z=1$.
The left panel is for the equal mass binary, and the right panel is for
 the unequal mass binary with the fixed companion masses
 $M_1=10^3, 10^4$ and $10^5 M_\odot$.  
We assume that the event rate $\Delta N$ for the various mass binaries
 $(10^{-1}-10^{5} M_{\odot})$ is the same as for the NS/NS
 binary for simplicity.
In the left panel, the errors are the smallest at a few $M_{\odot}$ mass, 
 because the frequency at $1-10$ yrs before the coalescence is 
 near $0.1$ Hz for these mass range and DECIGO/BBO is the most
 sensitive there. 
For a larger mass binary $M_{1,2} \gsim 10^2 M_{\odot}$, though
 $S/N$ is larger than that for NS/NS binary ($1.4+1.4 M_\odot$), the
 frequency is shifted to the lower one ($< 0.1$ Hz) so that
 the errors become large because DECIGO/BBO is less sensitive there.    
In the right panel, the observational period is fixed with
 $\Delta t=3$ yr but the result simply scales as
 $(\Delta t/3 {\text{yr}})^{-2}$ from Eq.(\ref{delx}).
For $M_1=10^3$ and $10^4 M_\odot$, the errors are small for the
 smaller companion masses, and the reason is the same as in the left panel.
For $M_1=10^5 M_\odot$ with the smaller companion mass
 ($M_2 \lsim 10 M_\odot$), the S/N is small and hence the error is large.

Finally, we discuss the estimation errors of the parameters of the dark
 energy by the ultimate DECIGO.
The dark energy is usually characterized by its equation of state
 $w(z)=P_w(z)/\rho_w(z)$ where $P_w(z)$ is the pressure and $\rho_w(z)$
 is the  energy density. 
In the cosmology with the dark energy, the Hubble parameter at $z$ is
 written as
\beq
  H(z) = H_0 \left[ \Omega_m (1+z)^3 + \left( 1-\Omega_m-\Omega_w \right)
 (1+z)^2 + \Omega_w \exp \left\{ 3 \int_0^z d z^{\prime}
 \frac{1+w(z^{\prime})} {1+z^{\prime}} \right\} \right]^{1/2},
\label{dehub}
\eeq
where $\Omega_w$ is the density parameter of the dark energy. 
Note here that no assumption on the curvature of the universe is made.
We use the conventional first order expansion of $w(z)$ as
 $w(z)=w_0+w_1 z$ where $w_0$ and $w_1$ are constants. 
Then the exponential in Eq.(\ref{dehub}) is rewritten as
 $(1+z)^{3 (1+w_0-w_1)} e^{3 w_1 z}$.
We calculate the estimation errors of four parameters
 ($\Omega_m,\Omega_w,w_0,w_1$) instead of the acceleration parameter
 $X(z)$ by inserting Eq.(\ref{dehub}) into Eq.(\ref{xz}).
We use the signal from all the NS/NS binaries at $z=0-2$ in
 Fig.\ref{fdn2}.

\begin{table}
\begin{center}
  \begin{tabular}{lcccc} \hline \hline
  \multicolumn{1}{c} {obs. period}~ & ~$\Delta \Omega_m / \Omega_m$~ &
    ~$\Delta \Omega_w / \Omega_w$~ & ~$\Delta w_0$~ &
    ~$\Delta w_1$ \\ \hline
  \multicolumn{2}{l} {(a) $w(z)=w_0$} & & & \\ 
  \multicolumn{1}{c} {$1$ yr} & $0.030$~ & ~$0.038$~ & ~$0.031$~ &
     --- \\
  \multicolumn{1}{c} {$3$ yr} & $4.2 \times 10^{-3}$~ &
     ~$5.3 \times 10^{-3}$~ & ~$4.3 \times 10^{-3}$~ & --- \\
  \multicolumn{1}{c} {$10$ yr} & $6.3 \times 10^{-4}$~ &
     ~$7.6 \times 10^{-4}$~ & ~$5.9 \times 10^{-4}$~ & --- \\
  \multicolumn{2}{l} {(b) $w(z)=w_0+w_1 z$} & & & \\
  \multicolumn{1}{c} {$1$ yr} & $0.21$~ & ~$0.62$~ & ~$0.37$~ &
    ~$0.81$ \\
  \multicolumn{1}{c} {$3$ yr} & $0.029$~ & ~$0.086$~ & ~$0.051$~ &
    ~$0.11$ \\
  \multicolumn{1}{c} {$10$ yr} & $4.2 \times 10^{-3}$~ &
     ~$0.012$~ & ~$7.3 \times 10^{-3}$~ & ~$0.016$ \\  \hline
  \end{tabular}
\end{center}
\caption{\label{t1} The estimation errors of cosmological parameters
 $(\Omega_m,\Omega_w,w_0,w_1)$ for various observational period
 ($1,3,10$ yr). The upper part is for (a) a constant $w$ (i.e. $w=w_0$),
 and the lower part is for (b) a variable $w$ (i.e. $w=w_0+w_1 z$).
 We use the sources of all the NS/NS binaries at $z=0-2$.}
\end{table}

\begin{table}
\begin{center}
  \begin{tabular}{lccc} \hline \hline
  \multicolumn{1}{c} {obs. period}~ & ~$\Delta \Omega_w / \Omega_w$~ &
 ~$\Delta w_0$~ & ~$\Delta w_1$ \\ \hline
  \multicolumn{2}{l} {(a) $w(z)=w_0$} &  & \\ 
  \multicolumn{1}{c} {$1$ yr} & ~$1.4 \times 10^{-3}$~ &
 ~$5.8 \times 10^{-3}$~ &  --- \\
  \multicolumn{1}{c} {$3$ yr} &
     ~$1.9 \times 10^{-4}$~ & ~$8.0 \times 10^{-4}$~ & --- \\
  \multicolumn{1}{c} {$10$ yr} & 
     ~$2.9 \times 10^{-5}$~ & ~$1.1 \times 10^{-4}$~ & --- \\
  \multicolumn{2}{l} {(b) $w(z)=w_0+w_1 z$} & &  \\
  \multicolumn{1}{c} {$1$ yr} & $6.6 \times 10^{-3} $~ &
 ~$7.3 \times 10^{-3}$~ & ~$0.058$~ \\
  \multicolumn{1}{c} {$3$ yr} & $9.4 \times 10^{-4}$~ &
 ~$9.6 \times 10^{-4}$~ & ~$8.0 \times 10^{-3}$~ \\
  \multicolumn{1}{c} {$10$ yr} & $1.4 \times 10^{-4}$~ &
 ~$1.2 \times 10^{-4}$~ & ~$1.1 \times 10^{-3}$~ \\  \hline
  \end{tabular}
\end{center}
\caption{\label{t2} Same as Table\ref{t1}, but for flat universe model
 ($\Omega_m+\Omega_w=1$).}
\end{table}

In Table \ref{t1}, the results are shown for various observational
 periods i.e., $1,3$ and $10$ yrs.
The upper part is for (a) a constant $w$ (i.e. $w=w_0$), and the
 lower part is for (b) a variable $w$ (i.e. $w=w_0+w_1 z$).
The results are shown for the cosmological parameters
 $(\Omega_M,\Omega_w,w_0,w_1)=(0.3,0.7,-1,0)$.
From the upper part of Table \ref{t1}, these parameters
 $(\Omega_m,\Omega_w,w_0)$ are determined within a few percent by
 $1$ yr observation, which is comparable to the expected accuracy
 from SNAP satellite \cite{snap04}.
We found that for the sources of small redshift $(z \lsim 0.5)$
 the parameters are not well determined because these parameters
 are degenerate.
But including the sources of high redshift ($z \gsim 1$), the degeneracy
 is resolved and these cosmological parameters are well determined.

Table \ref{t2} is the same as Table \ref{t1}, but assuming the flat universe
 model ($\Omega_m+\Omega_w=1$) suggested from the cosmic microwave background
 (CMB) observation by WMAP \cite{s03}.
The results are about $10$ times better than that in Table \ref{t1}.

We comment on the results for practical DECIGO or BBO.
The errors by using these detectors are $1000$ times larger than
 the results in the previous figures (Figs.\ref{dx18p} to \ref{dx26p})
 and tables (Table \ref{t1} and \ref{t2}).
But since the errors are proportional to $\Delta t^{-2}$ ($\Delta t$ is
 the observational period) from Eq.(\ref{delx}), the cosmic acceleration
 could be measured with large observational period $\Delta t \gsim 10$ yr.
In fact, the parameters $(\Omega_m,w_0)$ will be determined within
 $(3 \%, 10 \%)$ by $10$ yr observation of practical DECIGO/BBO 
 from Table \ref{t2} (a).

\section{Conclusion}

We have discussed the possibility of the direct measurement of
 the cosmic acceleration by using the decihertz gravitational
 wave detector.
We take the binary of mass $0.1-10^5 M_{\odot}$ at redshift $z=0-2$
 as the source measured by the ultimate DECIGO.
Following the previous paper \cite{skn01}, we newly include the effects
 of (1) the number of the sources and (2) the various masses and redshifts
 of the sources.
We found that the expansion rate of the Universe (i.e. Hubble parameter)
 at $z=0-2$ is directly determined within $\sim 10 \% ~(0.1 \%)$
 for $1 ~(10)$ yr observation by the  ultimate DECIGO.
The equation of state of the dark energy $w$ is determined within $\sim
 3 \% ~(0.06 \%)$ for $1 ~(10)$ yr observation. 
Even for practical DECIGO or BBO, $w$ will be determined within $\sim 10\%$
 by $10$ yr observation assuming the flat universe model. 
Hence the DECIGO or BBO will determine the nature of the dark energy
 with high accuracy independently from the other cosmological tests
 such as SNAP, CMB observations.

\section*{Acknowledgements}
We would like to thank Peter Bender, Bernard Schutz, and
 Naoshi Sugiyama for useful comment and discussions.
This work was supported in part by
a Grant-in-Aid for the 21st Century COE
``Center for Diversity and Universality in Physics''
and also supported by Grant-in-Aid for Scientific Research
of the Japanese Ministry of Education, Culture, Sports, Science
and Technology,
No.14047212 (TN), and No.14204024 (TN).

%

\end{document}